\documentclass[letter]{aa} 

\usepackage{graphicx}
\usepackage{txfonts}
\usepackage{amsmath}
\usepackage{gensymb}
\usepackage{comment}

\mathchardef\mhyphen="2D

\def\j0501{eRASS1\,J050129.5-073309\xspace} 

\begin{document}

   \title{Discovery of the lensed quasar \j0501 with SRG/eROSITA and {\it Gaia}}

   \author{Dusán Tubín-Arenas\thanks{\email{dtubin@aip.de}}
          \inst{1,2}
         \and
         Georg Lamer\inst{1}
        \and
         Mirko Krumpe\inst{1}
         \and
         Tanya Urrutia\inst{1}
         \and
             Axel Schwope\inst{1}
             \and
             Roisín Brogan\inst{1}
             \and
             Johan Comparat\inst{3}
             \and
             Mara Salvato\inst{3}
             \and
             Esra Bulbul\inst{3}
             \and
             Christian Garrel\inst{3}
             \and
             Malte Schramm\inst{4}
             \and 
             Teng Liu\inst{3}
}

   \institute{Leibniz-Institut f\"ur Astrophysik Potsdam (AIP), An der Sternwarte 16, 14482 Potsdam, Germany
         \and
    Potsdam University, Institute for Physics and Astronomy, Karl-Liebknecht-Straße 24/25, 14476 Potsdam, Germany 
\and
         Max-Planck-Institut f\"ur extraterrestrische Physik, Gießenbachstraße 1, 85748 Garching, Germany
        \and
        Saitama University, 255 Shimookubo, Sakura Ward, Saitama, Japan}

   \date{Received ...; accepted ...}

  \abstract
   {We report the discovery and spectroscopic identification of the bright doubly lensed quasar \j0501 at redshift $z=2.47$, selected from the first all-sky survey of the Spectrum Roentgen Gamma (SRG) eROSITA telescope and the {\it Gaia} EDR3 catalog. We systematically searched for extragalactic sources with eROSITA X-ray positions that have multiple {\it Gaia} counterparts, and we have started spectroscopic follow-up of the most promising candidates using long-slit spectroscopy with NTT/EFOSC2 to confirm the lens nature.
The two images are separated by 2.7\arcsec, and their average {\it Gaia} {\it g}-band magnitudes are 16.95 and 17.33. Legacy Survey DR10 imaging and image modeling reveal both the lensing galaxy and tentatively the lensed image of the quasar host galaxy. Archival optical light curves show evidence of a variability time delay, with the fainter component lagging the brighter by about 100 days. The brightness of the fainter image has also decreased by about one magnitude since 2019. This dimming was still obvious at the time of the spectroscopic observations and is probably caused by microlensing. The optical spectroscopic follow-up obtained from NTT/EFOSC2 and the evidence provided by the imaging and timing analysis allow us to confirm the lensed nature of \j0501.
   }
\authorrunning{D. Tubín-Arenas et al.}

   \maketitle
%

\section{Introduction}

The alignment between a distant quasar and a massive foreground galaxy can produce multiple images of the background quasar due to the space and time distortion produced by the gravitational potential of the massive galaxy. Gravitationally lensed quasars are crucial to understanding different aspects of galaxy evolution and cosmology, such as the baryonic and dark mass distribution of the lens \citep{Bate2011,10.1093/mnras/stu106}, the host galaxy properties of the lensed quasar \citep[e.g.,][]{Peng_2006}, and the structure and accretion properties of the quasar \citep{refId0,2021A&A...654A.155H,2022A&A...659A..21P}. 

The Hubble constant, $H_{0}$, can be constrained through time-delay cosmography by studying the time lag between the lensed images of the quasar and the mass distribution of the lens \citep{10.1093/mnras/132.1.101,10.1093/mnras/stw3006,2017MNRAS.465.4895W,2020MNRAS.494.6072S,2020MNRAS.498.1420W}. The use of time-delay measurements in lensed quasars is still limited by the scarcity of suitable objects since only a small fraction of the few hundred currently known quasar lenses can be used for such studies.

Currently underway are several studies that use the resolving power of {\it Gaia} to identify lensed systems from various input quasar samples selected optically or from the wide-field Infrared Survey Explorer (WISE) catalogs \citep{Agnello2017,Delchambre2019,Lemon2023}. Optical- and near-infrared-selected quasar samples for lens searches are affected by the blending of the quasar images with the emission from the lensing galaxies. This potentially leads to an incompleteness of the input samples.
In X-rays, the emission of the lensing galaxy can generally be neglected, and since most bright extragalactic X-ray point sources are quasars or other active galactic nuclei, X-ray catalogs are an ideal input sample for lens searches. The extended ROentgen Survey with an Imaging Telescope Array \citep[eROSITA;][]{2012arXiv1209.3114M,2021A&A...647A...1P} All-Sky Surveys (eRASS) have dramatically increased the number of X-ray-selected quasars, to several $10^6$ objects. With the completion of the first eROSITA All-Sky Survey \citep[eRASS1;][]{Merloni2023}, which lasted for six months, we initiated a project to identify the lensed quasars in the resulting catalog of X-ray sources.

In this Letter we report the discovery of a new gravitational-lensed quasar and the spectroscopic confirmation of the lens nature of \j0501 with eROSITA, {\it Gaia}, and New Technology Telescope (NTT)/ESO Faint Object Spectrograph and Camera (EFOSC2) observations. This Letter is structured as follows: In Sect. \ref{sec:data} we describe the method for selecting lensed quasar candidates, the observations, and the data reduction of the spectroscopic follow-up. The main results are shown in Sect. \ref{sec:results}, while the discussion and conclusions are presented in Sect. \ref{sec:conclusions}. Throughout this Letter we assume a flat $\Lambda$ cold dark matter cosmology with $h_{0}=0.677$ and $\Omega_{m}=0.31$ \citep{2020A&A...641A...6P}. We adopt the standard astronomical orientation with north up and east to the left. All measurement uncertainties given refer to 68\% confidence limits.
\section{Selection method, observation, and data reduction}\label{sec:data}
The lensed quasar candidates were selected based on the X-ray emission detected by eROSITA and the multiplicity of {\it Gaia} counterparts. The details of the selection and the analysis of the full sample will be presented in a future paper (Tubín et al. in prep.).

\subsection{Selection method}

We used the eRASS1 catalog as input for the identification of gravitational lens candidates. 
The eROSITA sky area with German data rights is delimited by Galactic longitudes $180 \degree < l < 360 \degree$.
For the ESO/NTT observations, we defined the survey area as the intersection of the German part of eRASS1, the equatorial coordinates $-30 \degree < \delta  < 15 \degree$, $0\,\mathrm{h} < \alpha < 15\,\mathrm{h}$, and galactic latitudes $ |b| > 15 \degree$. This results in a survey area of about $6500\, \mathrm{deg}^2$. Within this survey area, we searched for multiple {\it Gaia} Early Data Release 3 \citep{2021A&A...649A...2L} counterparts within 12\arcsec~of the X-ray positions. We selected for separations lower than 5\arcsec\ between the multiple {\it Gaia} counterparts.  

We removed obvious galactic stars by using the {\it Gaia} parallax and proper motion parameters. To exclude stellar coronal emitters, we applied a cut in the optical to X-ray spectral energy distribution according to \citet{Maccacaro1988}:

\[ \mathrm{log}(f_\mathrm{x}/f_\mathrm{opt}) = \log_{10} (f_{0.2-2.3 \,\mathrm{keV}}) +  m_\mathrm{G}/2.5 + 5.37   > -1.5. \]

The cut in $f_\mathrm{x}/f_\mathrm{opt}$ removed about $10\%$ of objects for which no parallax or proper motion information is available. Applying this additional cut would not remove any known lenses from the Cambridge Gravitationally Lensed Quasar (GLQ) database\footnote{\url{https://research.ast.cam.ac.uk/lensedquasars/}} \citep{10.1093/mnras/sty3366}.

We performed a visual inspection using Panoramic Survey Telescope and Rapid Response System (Pan-STARRS) \citep{2016arXiv161205560C} and Dark Energy Survey (DES) \citep{2021ApJS..255...20A} imaging to clean the resulting sample from contaminations, such as large galaxies, galaxy-galaxy pairs, quasar-galaxy pairs, and galactic stars. 
With this method, we found 115 lensing candidates in the survey area. Cross-matching this set of candidates with SIMBAD, the GLQ database, and the ESO archive shows that about half (58) of these objects are already known quasar lenses or candidates.

\subsection{NTT/EFOSC2 observations}

The first 30 lensing candidates were spectroscopically observed in November 2022 with the ESO/La Silla 3.6 m  NTT/EFOSC2 under ESO program 110.248X. The spectroscopic observations were performed with grating \#6 and a 1\arcsec~slit width, which corresponds to a wavelength resolution of 15.5\,\AA~(full width half maximum) in the range 3860--8070\,\AA. The order-separating filter GG375 was used to reduce contamination by the second-order spectrum. Since all observed objects have two candidate lensing images, we rotated the slit using the {\it Gaia} coordinates in order to cover both images with one exposure.  

The object \j0501 was observed on November 23, 2022, with a spectroscopic exposure of 1800 seconds. The seeing at the time of the observation, as measured in the R-band acquisition image, was about 0.8$\arcsec$ (full width half maximum). With an image separation of $\sim 2.7\arcsec$, the two spectra are well separated. 

We followed the standard data reduction process using the ESO-MIDAS \citep{1992ASPC...25..115W} data reduction package to subtract the bias, correct for pixel-to-pixel variations (flat field), and perform the wavelength calibration with exposures of He and Ar lamps. The spectra were extracted using an algorithm that optimizes the signal-to-noise ratio and rejects cosmic ray hits (MIDAS {\tt SKYFIT/LONG} and {\tt EXTRACT/LONG}). For the flux calibration, spectra of the standard star Feige 110 \citep{2014A&A...568A...9M} were taken and reduced in the same way as the science spectra.

\begin{figure}
    \centering
    \includegraphics[width=\columnwidth]{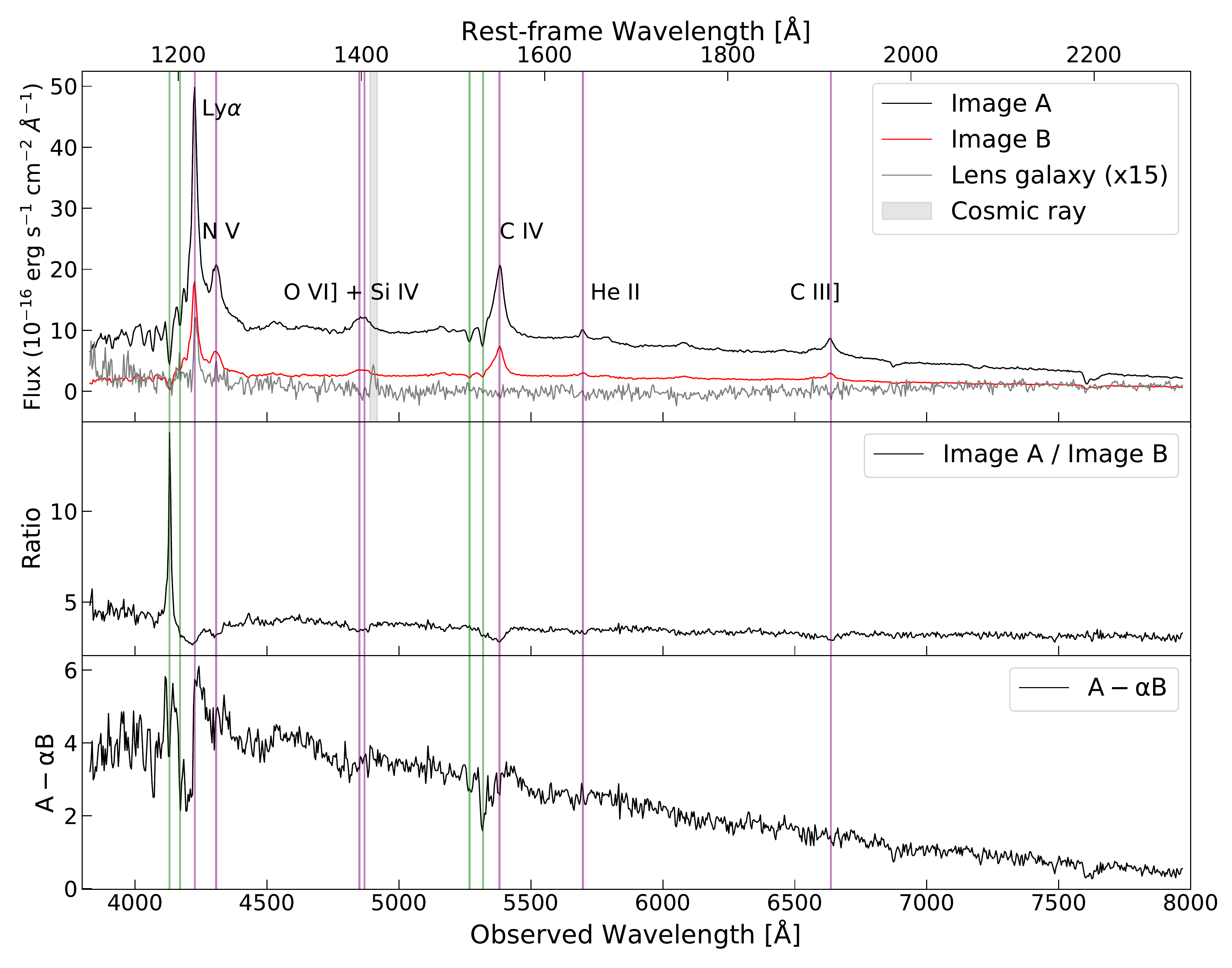}
    \caption{NTT/EFOSC2 optical spectra of the two components of the lensed quasar \j0501. {\it Top panel:} The black and red curves correspond to image A and image B of the doubly lensed quasar. The gray curve corresponds to the central Moffat profile fitting of the 2D spectrum. The flux of this spectrum is increased by a factor of 15 for visualization purposes. We show observed and rest-frame wavelengths to illustrate the redshift of the source. We annotate the main emission lines recognized in the spectra. {\it Middle panel:} Ratio between the spectra from image A and image B. {\it Bottom panel:} Subtraction of scaled spectrum B from spectrum A using a scaling factor of $\alpha=2.4$ (see Sect.~\ref{scaling_factor} for more details). The vertical purple and green lines highlight the main emission and absorption lines, respectively. The velocities of the absorption features are $-3460$ km s$^{-1}$ and $-6320$ km s$^{-1}$ with respect to $z=2.47$.}
    \label{fig:spec}
\end{figure}

\begin{figure}
\begin{minipage}[t]{0.48\textwidth}
    \includegraphics[width=\columnwidth]{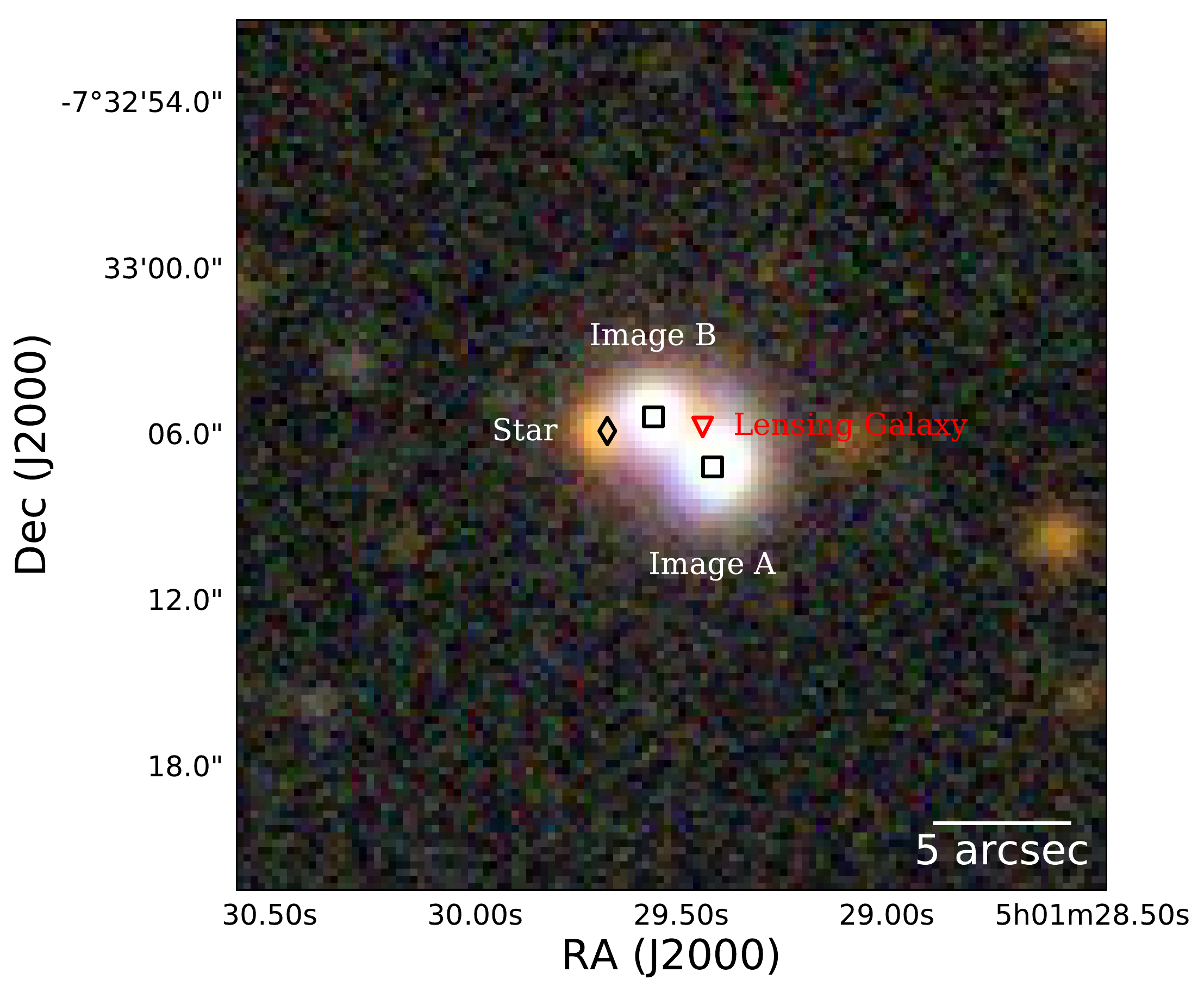}
\end{minipage}
\begin{minipage}[t]{0.48\textwidth}
    \includegraphics[width=\columnwidth]{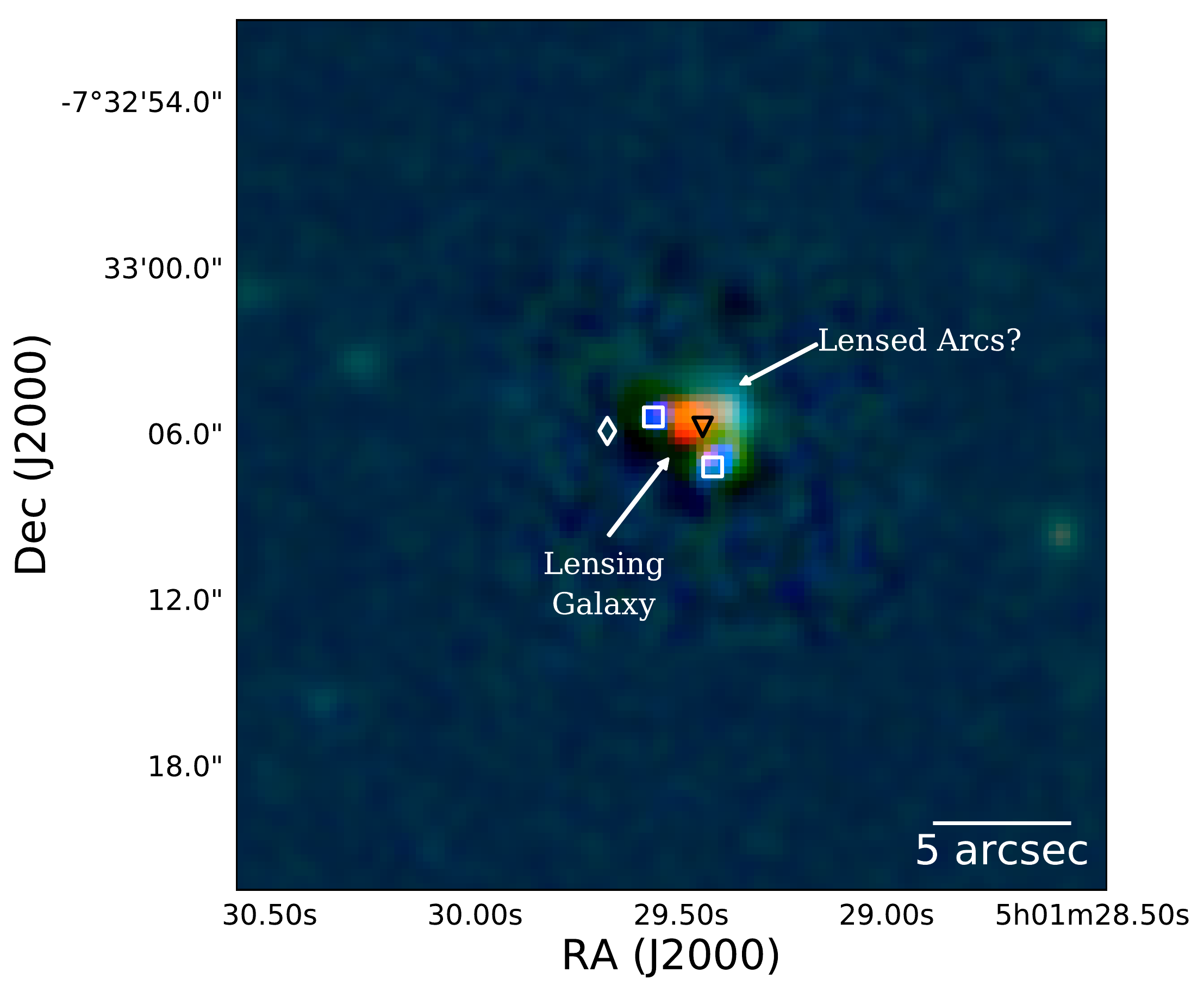}
\end{minipage}
\centering
\begin{minipage}[t]{0.48\textwidth}
    \includegraphics[width=\columnwidth]{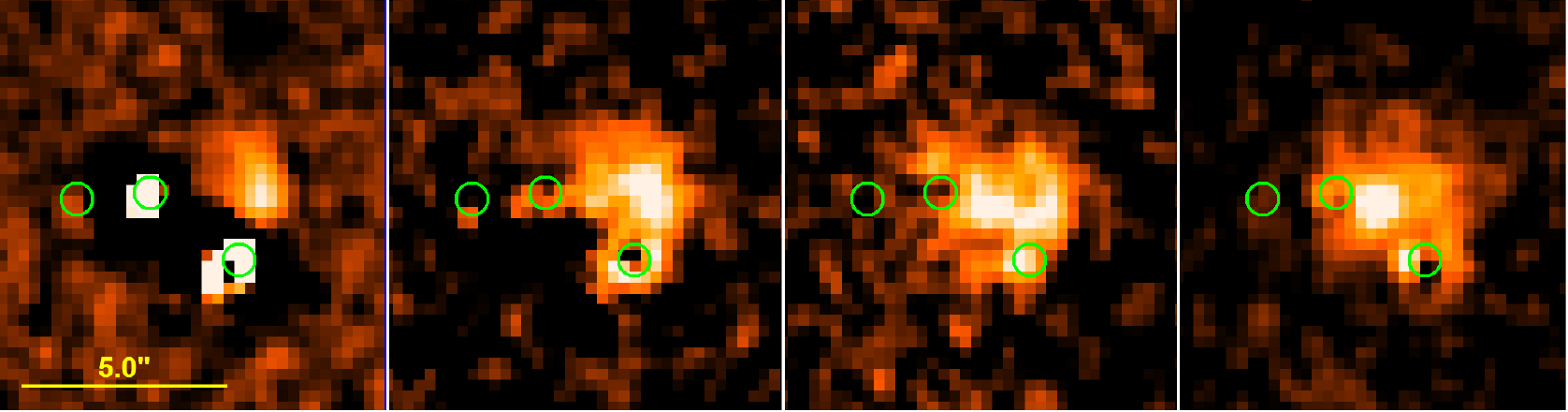}
\end{minipage}

\caption{Optical imaging of \j0501. {\it Top panel:} Combined {\it g-, r-}, and {\it z}-band LS DR10 (DECals) images centered on the source. {\it Middle panel:} Red-green-blue image of the residuals after subtracting three point sources (the star and the lensed images) with GALFIT. We mark the lensed images of the quasar with black (top) and white (bottom) squares. The {\it Gaia} position of the star is marked with a white (top) and black (bottom) diamond. The LS DR10 position of the lensing galaxy is marked with a red and black triangle in the top and bottom panels, respectively. We also annotate the position of the likely lensed arcs and the lensing galaxy. {\it Bottom panel:} Individual LS DR10 {\it g, r, i,} and {\it z} residual images of the lensing galaxy and arcs after point-source subtraction of the lensed quasar images and the foreground star (green circles). The scale bars have an angular size of 5\arcsec.}
\label{fig:image}
\end{figure}

\section{Results}\label{sec:results}

Here we present the results for \j0501, which is the optically brightest object in our sample and whose remarkable properties call for further follow-up observations, such as high-resolution imaging and photometric monitoring of its components.

\subsection{NTT/EFOSC2 spectroscopic data}
\label{scaling_factor}

Figure \ref{fig:spec} shows the NTT/EFOSC2 optical spectra of \j0501. We highlight the most prominent emission lines, including Ly$\alpha$, N V $\lambda1238$, O IV] $\lambda1399$, Si IV $\lambda1402$, C IV $\lambda1549$, He II $\lambda1640$, and C III $\lambda1908$, with vertical purple lines. Since the quasar is doubly lensed, we identify image A as the brighter lensed component and image B as the fainter (see Fig. \ref{fig:image}). We fit the C IV and He II emission lines with a Gaussian profile and derive a redshift of $z=2.47\pm0.03$ for both spectra. The conformity of the spectra and the presence of a candidate lens galaxy confirms the lensed nature of the source. 

The spectra of both components show absorption line features (green vertical lines); the most prominent ones are visible in the blue wings of the C IV and Ly$\alpha$ lines. We interpret these features as indications of outflowing material within the quasar system. For the two most prominent lines near the C IV emission line, we measure outflow velocities of -3460 km s$^{-1}$ and -6320 km s$^{-1}$. Absorption lines consistent with these velocities can also be found for Ly$\alpha$. 

In order to extract the spectrum of the lensing galaxy (see Fig. \ref{fig:image}), the much brighter quasar spectra need to be subtracted. Since the spatial profiles
of the spectra vary as a function of wavelength, we used the following method to de-blend the spectra: we first averaged the two-dimensional sky-subtracted spectrum in blocks of 50 wavelength pixels ($\sim$206\,$\AA$) and fit each of the averaged blocks with three identical Moffat \citep{1969A&A.....3..455M} spatial profiles. The positions of the three profiles were fixed using the positions for the QSO images and the lensing galaxy from Table \ref{tab:position}. 
The spatial parameters of the Moffat profile were then interpolated to each wavelength pixel, and the profiles were fitted at each wavelength with only the intensities $\rm I_{0,i}$ left free to vary. The resulting spectrum for the lensing galaxy is shown as a gray line in the top panel of Fig. \ref{fig:spec}. However, the signal-to-noise ratio of this spectrum is too low to determine a redshift for the lens. 
Subtracting the two Moffat profiles of the quasar images from the two-dimensional spectra also reveals spatially extended narrow $\mathrm{Ly}\alpha$ emission from the quasar host galaxy (see Fig. \ref{fig:spec}).

The ratio between the two lensed images (bottom panel of Fig. \ref{fig:spec}) shows that the continuum of image B is three to four times fainter and somewhat redder than that of image A. At the same time, the equivalent width of the emission lines is higher in image B. This could be due to microlensing affecting the lensed images, caused by stars in the lensing magnifying or demagnifying selected spatial regions of the quasar. Interestingly, on the blue side of Ly$\alpha$, the absorption feature attributed to the faster outflow is much more prominent in image B, as seen in the ratio spectrum. This ``glitch'' could be due to microlensing, intrinsic variability, or an intervening absorption system unrelated to the quasar. 

We plot a difference spectrum in the bottom panel of Fig. \ref{fig:spec}, which was derived by subtracting the spectrum of image B from spectrum A with the following scaling:
$\rm A - \alpha B$, where $\alpha$ is optimized to remove the
emission lines Ly$\alpha$, \mbox{C IV $\lambda1549$}, and C III $\lambda1908$. The resulting scaling factor of $\alpha=2.4$ was obtained by averaging the ratio of the integrated flux values of the three corresponding emission lines in both spectra. This factor can be interpreted as the ratio between the macro-magnification factors of the continuum and emission lines, assuming that the lines are not affected by microlensing.

\subsection{Optical and near-infrared imaging }

\begin{table*}[htbp]
    \caption{Position of the \j0501 components. The quasar images and the foreground star are detected by both {\it Gaia} and LS DR10, while the lensing galaxy is only detected by LS DR10. We also list the relative position ($\Delta \alpha,\Delta \delta$) of the sources in the field of view with respect to the {\it Gaia} position of image A.}
    \label{tab:position}
    \centering
    \begin{tabular}{llllllllll}
       \hline
       Component & Telescope & $\alpha$(2000.0)& err $\alpha$ & $\delta$(2000.0)& err $\delta$&$\Delta \alpha$&  err $\Delta \alpha$& $\Delta \delta$ &err $\Delta \delta$ \\
            &  & [deg] & [mas]     &   [deg]& [mas]  & [\arcsec]& [mas]& [\arcsec]& [mas] \\
       \hline
       Image A  & {\it Gaia} &  $75.372574$& $ 0.05$ & $-7.551979$& $ 0.04$ &--- &---&---&---\\
       Image B  & {\it Gaia} & $75.373176$& $ 0.06$ & $-7.551526$& $ 0.05$ & $2.17$ &$0.08$ & $1.63$&$0.06$   \\
       Lensing galaxy  & DECals DR10 &$75.3727$& $ 11.89$ & $-7.5516$&$ 11.11$  & $0.38$&$11.89$ &$1.49$& $11.11$\\  
       Foreground star & {\it Gaia} &   $75.373665$&$ 1.58$& $-7.551639$ &$ 1.11$& $3.93$&$1.58$  &$1.22$&$1.11$\\  
       \hline
    \end{tabular}
\end{table*}

Legacy Survey Data Release 10 \citep[LS DR10;][]{2019AJ....157..168D} imaging from Fig. \ref{fig:image} shows the lensed images (image A and image B) of \j0501 with a separation of 2.7\arcsec. Recently, \j0501 was selected as a lensed quasar candidate by \cite{2023arXiv230111080H}, based on these Legacy Survey images. Within the field of view, it is possible to note a foreground star, whose position is highlighted by a diamond. The star has a mean {\it g}-band magnitude of $20.84$ and a significant proper motion ($\mu_\mathrm{\alpha}=4.4 \pm 1.9\, \mathrm{mas\, yr^{-1}},\, \mu_\mathrm{\delta} = -13.2 \pm 1.7 \, \mathrm{mas\, yr^{-1}}$).

The LS DR10 catalog also lists a potential lensing galaxy positioned between the two quasar images, which is highlighted with the red triangle. Table \ref{tab:position} summarizes the position of the objects within the field of view of Fig. \ref{fig:image}. 

To confirm its presence and visualize the lensing galaxy, we used GALFIT \citep[][]{2002AJ....124..266P,2010AJ....139.2097P} to fit three point sources and the candidate lensing galaxy to LS DR10 $r$, $i$, and $z$ images. As a point spread function reference, we used a nearby star with {\it Gaia} Data Release 3 (DR3) source\_id=3187010525569573760. The galaxy was modeled with a S\'ersic profile. Since the S\'ersic index is not well constrained by the fits, we set it to $n=4.0$. The resulting parameters for the galaxy are listed in Table \ref{tab:galfit}.

\begin{table}[htbp]
\fontsize{8.9}{10}\selectfont
    \caption{Best-fit parameters for the candidate lensing galaxy from GALFIT modeling of LS DR10 images. The coordinates are relative to the {\it Gaia} coordinates of image A. The model is a radially symmetric S\'ersic profile with a fixed S\'ersic index $n=4.0$.}
    \label{tab:galfit}
    \centering
    \begin{tabular}{lllll}
       \hline
       band &  $\Delta \alpha$ & $\Delta \delta$ &  mag   & $R_\mathrm{eff}$ \\
            &   [\arcsec]           &   [\arcsec]          &        &  [\arcsec]  \\
       \hline
       $r$   &  $-0.38\pm 0.03$ & $1.55\pm0.04$ &  $20.71 \pm  0.07$ &  $1.06 \pm 0.02$ \\
       $i$   &  $0.38\pm 0.03$ &  $1.55\pm 0.03$ &  $19.22 \pm  0.06$ &  $2.80 \pm 0.23 $ \\
       $z$   &  $0.63\pm0.03 $ &  $1.55\pm0.02$ &  $18.55 \pm 0.03 $ &  $3.11 \pm 0.15  $ \\  
       \hline
    \end{tabular}
\end{table}

The $g$-band image, where the lensing galaxy is not detected via GALFIT, was modeled with only three point sources. Subtracting the point source models from the $g,r,z$ images
results in the residual map shown in Fig. \ref{fig:image}. In addition to the lensing galaxy, an arc-shaped feature is apparent in the residual map. This feature is visible
in all bands (see the bottom panel Fig. \ref{fig:image}) and can be interpreted as a magnified image of the quasar host galaxy.
However, observations with high spatial resolution are required to confirm the detection of the quasar host galaxy.

\begin{figure*}
    \centering
    \includegraphics[width=\textwidth]{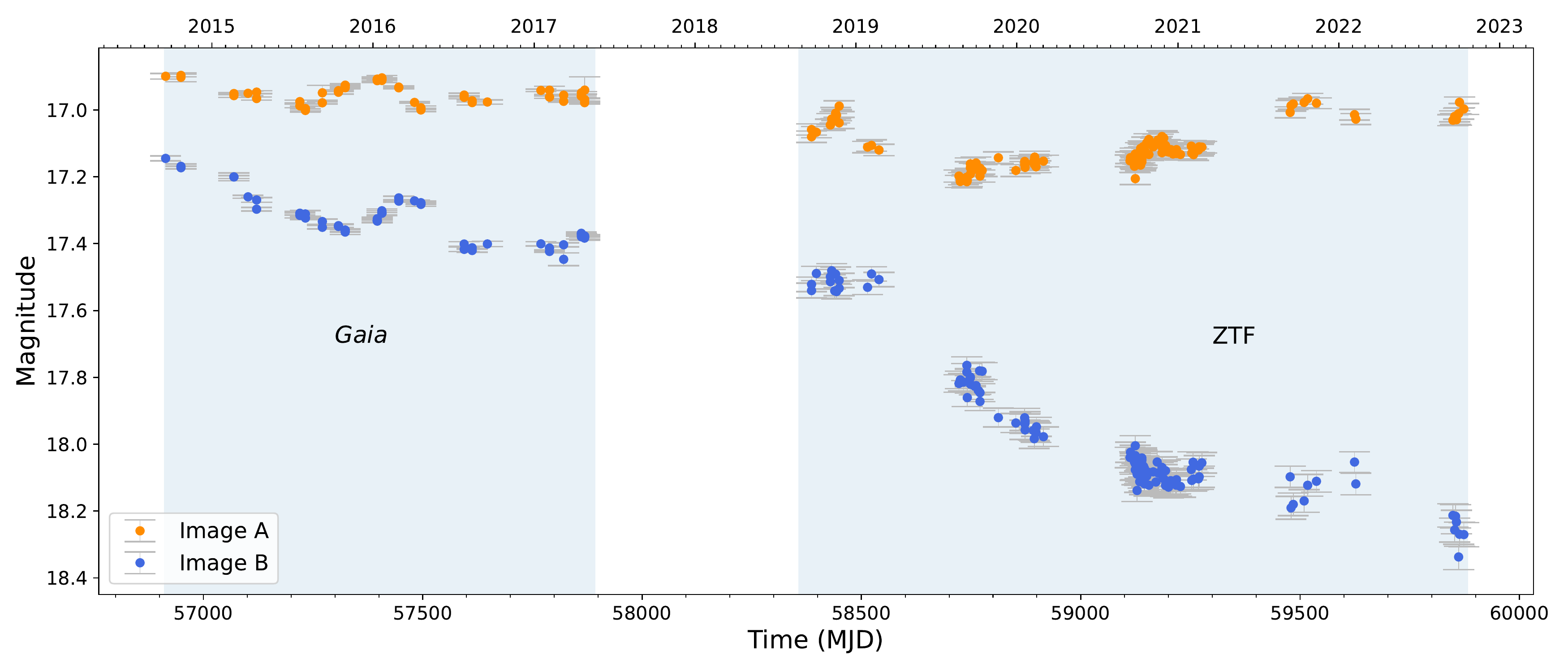}
    \caption{{\it Gaia} DR3 and ZTF {\it g}-band light curves for \j0501. The orange and blue circles correspond to image A and image B, respectively. The {\it Gaia} data cover the epoch from September 2014 to April 2017 and the ZTF data the epoch from October 2018 to August 2022.}
    \label{fig:lc}
\end{figure*}

\begin{figure}[h]
    \centering
    \includegraphics[width=\columnwidth]{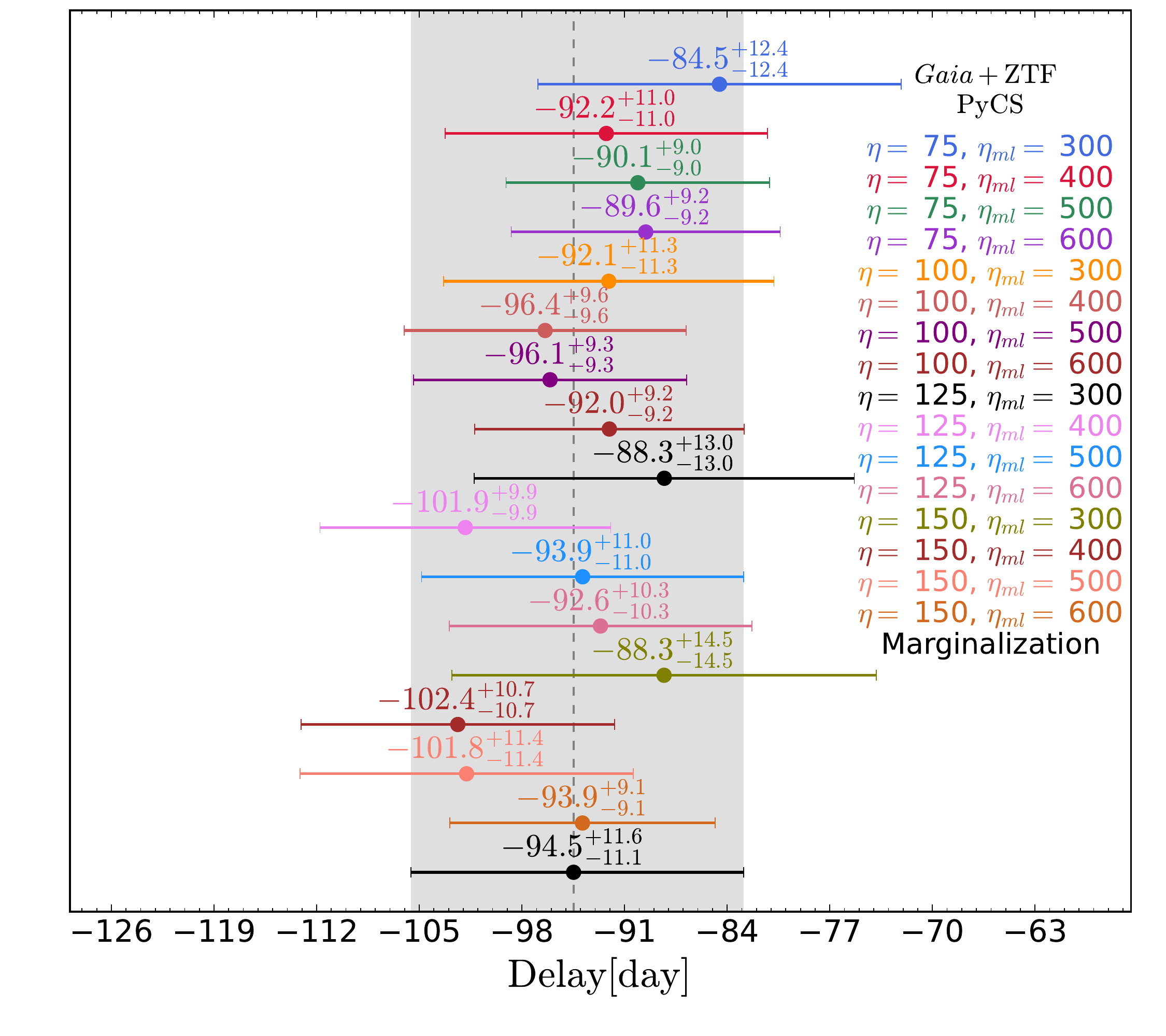}
    \caption{Time-delay measurements obtained from the free-knot spline fitting method. Each time delay was calculated with a different set of $\eta$ and $\eta_{ml}$. The black measurement at the bottom is the combined marginalization, which is shown as the gray shaded area.}
    \label{fig:splinefit}
\end{figure}

\begin{figure*}[h]
    \centering
    \includegraphics[width=\textwidth, height=\textwidth]{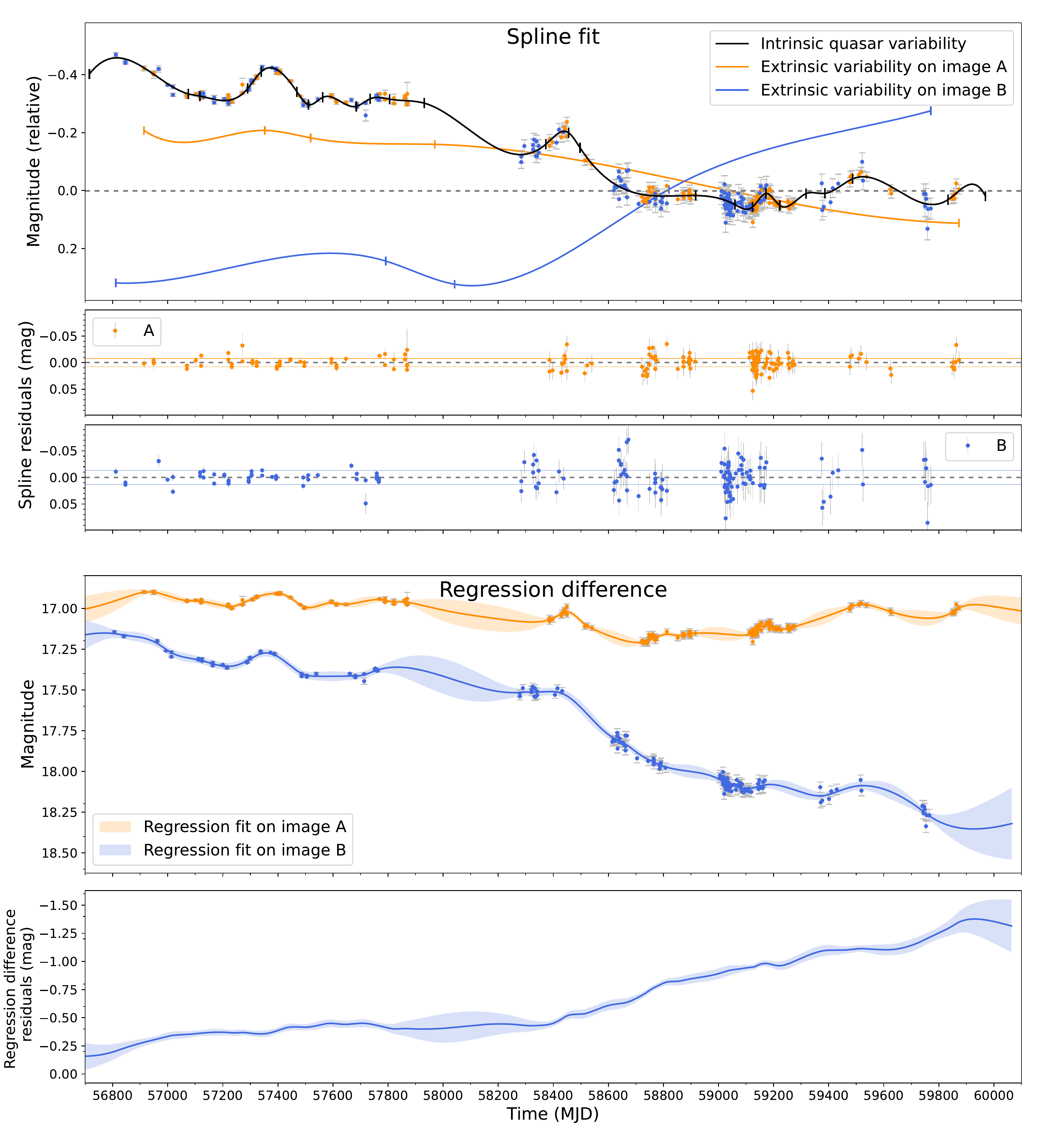}
    \caption{\texttt{PyCS3} fitting of the \textit{Gaia} + ZTF light curve. \textit{Top panels:} Spline fit and residuals of the \j0501 data. The solid black curve corresponds to the intrinsic variability of the quasar shared by the lensed images. The solid orange and blue lines show the extrinsic variation produced by microlensing affecting images A and B, respectively. The light curve is fitted with spline functions with $\eta=125$ and $\eta_{ml}=500$ days. \textit{Bottom panels:} Regression difference fit of the light curves and the corresponding 1$\sigma$ errors. The bottom panel shows the difference between the best fit of image A and the best fit of the time-shifted image B, evidencing microlensing effects. In the top four panels, the light curves of image B (blue data points) are shifted by the optimal time delay found with the corresponding method. For the case of the spline fitting, we apply a shift of $-94$ days, and for the regression difference method a shift of $-108$ days. The data points for image A (orange) are plotted without a time shift. }
    \label{fig:pycs3}
\end{figure*}

\subsection{Timing analysis }

Figure \ref{fig:lc} presents the optical light curves of the lensed quasar from {\it Gaia} DR3 \citep{2021A&A...649A...1G} and the Zwicky Transient Facility \citep[ZTF;][]{Masci_2019} data.  

Time-resolved {\it Gaia} photometric data for both lensed images are available in the epoch between September 2014 and April 2017 and do not overlap with the ZTF light curve. In the {\it Gaia} data, a bump-like feature is present in image A around day 57400. This feature in the light curve seems to be repeated approximately three months later in image B.

We note that ZTF can marginally resolve both lensed images of the quasar. To reduce the cross-contamination between the lensed images, we only considered the data taken when seeing was better than 2\arcsec. The ZTF light curve shows that the second image has decreased its flux by almost one magnitude since 2019 ($\sim58400$ MJD), while the primary image has stayed rather constant. Typical time lags of lensed quasars range from days up to a few months \citep[see Table 3 of ][]{2020A&A...640A.105M}. In this case, the brightness of image B has decreased for almost four years, which is typical for microlensing de-magnification caused by stars in the lensing galaxy; similar examples can be found in \citet{2020A&A...640A.105M}. 

The repeated bump in the {\it Gaia} light curve suggests a time delay between image A and image B. In order to quantify the tentative time lag, we fit the light curves with \texttt{PyCS3}\footnote{\url{https://gitlab.com/cosmograil/PyCS3}} \citep[][]{Millon2020}, a toolbox developed by the COSMOGRAIL\footnote{\url{https://www.cosmograil.org}} collaboration to estimate time delays between multiple images of strongly lensed quasars. 

\texttt{PyCS3} is a curve-shifting technique that implements two estimators to optimize the time delay of a set of light curves. The original formalism is extensively described in \cite{2013A&A...553A.120T}, and the automated version of PyCS3 applied to a large monitoring data set is presented in \cite{2020A&A...640A.105M}. \texttt{PyCS3} uses two estimators. The free-knot spline estimator fits a unique free-knot spline function to model the global intrinsic variations in the quasar and individual splines to model the extrinsic variations due to microlensing. 
The regression difference estimator applies Gaussian process regression independently to each light curve. The regressions are shifted in time and subtracted pair-wise. The algorithm minimizes the variability of the difference between the curves by varying the time shift \citep[see][ for more details]{2013A&A...553A.120T}. 

The main parameters of the spline estimator are the initial knot steps $\eta$ and $\eta_{ml}$, which correspond to the mean spacing between knots before starting the optimization of the intrinsic and extrinsic splines, respectively. We considered a set of values for different $\eta$ and $\eta_{ml}$ in order to find the time delay between the light curve using the free-knot spline method. The different combinations are listed in Fig. \ref{fig:splinefit}. We note that small values of $\eta$ and $\eta_{ml}$, namely $\eta<75$ and $\eta_{ml}<300$, reduce the spacing between the initial knots, resulting in an overfitting of the light curve, especially in the gaps where there are no data. The parameters for the regression difference method are the covariance function, amplitude, characteristic timescale, and observation variance. We set the amplitude parameter to 2.0 magnitudes, a scale of 200 days, a Matérn covariance function, and a smoothness degree $\nu = 1.5$. 

To estimate uncertainties on the time delay, we followed the approach described in \cite{2013A&A...553A.120T} and \cite{2020A&A...640A.105M}. \texttt{PyCS3} applies the curve-shifting techniques to a large number of synthetic mock light curves derived from the data with known time delays. These curves are drawn from a light curve model that mimics both the observed intrinsic and extrinsic variability of the real observations while controlling the hidden parameters, such as the time delay. We created 1000 mock light curves to derive uncertainties for the time delay.

We find that both estimators give consistent time delays between image A and image B. The free-knot spline estimator gives a time delay of $\rm \Delta T_{AB}=-94.5\pm 11$ days, and the regression difference estimator a time delay of $\rm \Delta T_{AB}=-107.8\pm 8.5$ days. 

Figure \ref{fig:pycs3} shows an example of the {\it Gaia} DR3 + ZTF light curve fitted with \texttt{PyCS3} estimators. The light curve is fitted with spline functions where $\eta=125$ and $\eta_{ml}=500$ days (top panels) and with the regression difference method (bottom panels). Each light curve of image B is shifted by 94 days in the case of the spline fitting, and 108 days for the regression difference method. We note that both estimators provide a direct way to highlight the microlensing effect over the quasar images. The spline fitting gives an example of the microlensing model, which affects the individual light curves. At the same time, the regression difference residuals show the difference between the time-shifted light curves, highlighting the overall microlensing effect on the quasar images. \j0501 is highly affected by microlensing, with an overall variation of $\sim 1.25$ magnitudes over $\sim 8$ years of archival data.
The reported time delay and the microlensing variability make \j0501 a suitable candidate on which to perform time-delay cosmography and study a likely caustic-crossing event, respectively. The transit of a high-magnification caustic due to an individual star, or caustic-crossing event, can magnify emission near the supermassive black hole (SMBH) with resolution on the scale of the SMBH event horizon, thus mapping the quasar structure \citep[see, e.g.,][]{2008A&A...480..327A,2015ApJ...798..138M}.  

\subsection{Spectrum Roentgen Gamma/eROSITA X-ray spectrum}

The X-ray emission of the lensed quasar cannot be resolved by eROSITA and can be treated as a point source. We extracted the X-ray spectrum using the eROSITA Science Analysis Software System \citep[eSASS, version eSASSuser211214; described in][]{2022A&A...661A...1B} task \texttt{srctool}. For the X-ray analysis, we only considered events in the 0.2--2.3 keV energy band stacked over the data from the first four eRASS catalogs (eRASS:4). We extracted the spectrum from a source region centered on the object with a radius of 1\arcmin~and a background annulus region with an inner and outer radius of 2.5 and 4.5\arcmin, respectively. 
We used the X-ray Spectral Fitting (XSPEC) package \citep{1996ASPC..101...17A} to fit the spectrum with an absorbed power law model (\texttt{tbabs*pow, abundance=wilm}). The spectrum was fitted using C-statistics \citep{1979ApJ...228..939C}.  
When fixing the column density at the galactic value of $4.55\times10^{20}$ cm$^{-2}$ \citep{2016A&A...594A.116H}, we obtain a photon index of $\Gamma= 2.13_{-0.51}^{+0.50}$.

We derive a flux of $3.21^{+0.88}_{-0.51}\times 10^{-13}\; \rm erg\; s^{-1}\; cm^{-2}$ in the 0.2--2.3 keV energy band, consistent with the stacked catalog eRASS:4 flux measurement of $3.15\times 10^{-13}\; \rm erg\; s^{-1}\; cm^{-2}$. \j0501 is one of the X-ray-brightest of all known lensed quasars detected by eROSITA (Tubín et al. in prep.), making it suitable for further X-ray follow-up observations.
Assuming our best-fit spectral model, we obtain a galactic-absorption-corrected rest-frame X-ray luminosity of $ L_{2 - 10\; \rm keV}=1.5\times 10^{46}\; \rm erg\; s^{-1}$. This source is more luminous than most of the quasars at similar redshifts (e.g., \citealt{Singal_2022}), implying significant lensing magnification.

\section{Conclusions}\label{sec:conclusions}

We present the discovery of the new lensed quasar \j0501 at redshift $z=2.47$.
The optical spectroscopic follow-up obtained with the NTT/EFOSC2 and the evidence provided by the imaging and timing analysis allow us to confirm the lensed nature of the source. We notice that image B of the quasar is highly affected by microlensing effects, which are evident when we analyze the ratio of the spectra in Fig. \ref{fig:spec} as well as its long-term light curve. 

The imaging analysis suggests the presence of a lensing galaxy with $r=20.7$ and a tentative detection of a lensing arc that we attributed to the quasar host galaxy. The confirmation of the lensing arcs could constrain the gravitational potential of the lens and enable the investigation of the quasar host galaxy properties, such as the ionized gas content, the nature of the ionizing radiation, star formation, the SMBH mass, or even the stellar population parameters of a source whose properties will be hidden without the presence of the lensing galaxy.

Finally, the {\it Gaia} DR3 light curve shows evidence of a time delay of $\sim 100$ days, where image B lags the brighter component as a consequence of intrinsic quasar variability and the lensed nature of the source. We see that the quasar images are affected by microlensing effects; image B shows a flux decrease of almost one magnitude over the last four years.  

The brightness of the source, the separation of the lensed images, and the time delay found in the optical light curve make \j0501 a perfect laboratory for performing high cadence multiwavelength follow-up and cosmological studies. Despite strong microlensing effects, the time delay is well constrained by the available data. The delay of $\sim 100$ days allows complete monitoring programs by current and future facilities, such as the \textit{Vera C. Rubin} Observatory with the 10-year Legacy Survey of Space and Time (LSST) project \citep{2019ApJ...873..111I}. Such optical monitoring programs, together with high-resolution observations and an accurate mass model of the lensing galaxy, can lead to a precise estimation of the Hubble constant via time-delay cosmography.

\begin{acknowledgements}

We thank the referee Frederic Courbin for useful suggestions which helped to 
improve the manuscript significantly.
D.T. acknowledges support by the German DLR under contract FKZ 50 OR 2203. 
M.K. acknowledges support from DFG grant number KR3338/4-1. 
G.L. acknowledges support from the German DLR under contract 50 QR 2104.
This work is based on data from eROSITA, the soft X-ray instrument aboard SRG, a joint Russian-German science mission supported by the Russian Space Agency (Roskosmos), in the interests of the Russian Academy of Sciences represented by its Space Research Institute (IKI), and the Deutsches Zentrum f\"{u}r Luft- und Raumfahrt (DLR). The SRG spacecraft was built by Lavochkin Association (NPOL) and its subcontractors, and is operated by NPOL with support from the Max Planck Institute for Extraterrestrial Physics (MPE). The development and construction of the eROSITA X-ray instrument was led by MPE, with contributions from the Dr. Karl Remeis Observatory Bamberg \& ECAP (FAU Erlangen-Nuernberg), the University of Hamburg Observatory, the Leibniz Institute for Astrophysics Potsdam (AIP), and the Institute for Astronomy and Astrophysics of the University of T\"{u}bingen, with the support of DLR and the Max Planck Society. The Argelander Institute for Astronomy of the University of Bonn and the Ludwig Maximilians Universit\"{a}t Munich also participated in the science preparation for eROSITA. The eROSITA data shown here were processed using the eSASS software system developed by the German eROSITA consortium.
Based on observations collected at the European Organisation for Astronomical Research in the Southern Hemisphere under ESO program 110.248X.
This work has made use of data from the European Space Agency (ESA) mission
{\it Gaia} (\url{https://www.cosmos.esa.int/gaia}), processed by the {\it Gaia}
Data Processing and Analysis Consortium (DPAC,
\url{https://www.cosmos.esa.int/web/gaia/dpac/consortium}). Funding for the DPAC
has been provided by national institutions, in particular the institutions
participating in the {\it Gaia} Multilateral Agreement.
Based on observations obtained with the Samuel Oschin Telescope 48-inch and the 60-inch Telescope at the Palomar
Observatory as part of the Zwicky Transient Facility project. ZTF is supported by the National Science Foundation under Grant
No. AST-2034437 and a collaboration including Caltech, IPAC, the Weizmann Institute for Science, the Oskar Klein Center at
Stockholm University, the University of Maryland, Deutsches Elektronen-Synchrotron and Humboldt University, the TANGO
Consortium of Taiwan, the University of Wisconsin at Milwaukee, Trinity College Dublin, Lawrence Livermore National
Laboratories, and IN2P3, France. Operations are conducted by COO, IPAC, and UW.
The Legacy Surveys consist of three individual and complementary projects: the Dark Energy Camera Legacy Survey (DECaLS; Proposal ID \#2014B-0404; PIs: David Schlegel and Arjun Dey), the Beijing-Arizona Sky Survey (BASS; NOAO Prop. ID \#2015A-0801; PIs: Zhou Xu and Xiaohui Fan), and the Mayall z-band Legacy Survey (MzLS; Prop. ID \#2016A-0453; PI: Arjun Dey). DECaLS, BASS and MzLS together include data obtained, respectively, at the Blanco telescope, Cerro Tololo Inter-American Observatory, NSF’s NOIRLab; the Bok telescope, Steward Observatory, University of Arizona; and the Mayall telescope, Kitt Peak National Observatory, NOIRLab. Pipeline processing and analyses of the data were supported by NOIRLab and the Lawrence Berkeley National Laboratory (LBNL). The Legacy Surveys project is honored to be permitted to conduct astronomical research on Iolkam Du’ag (Kitt Peak), a mountain with particular significance to the Tohono O’odham Nation. NOIRLab is operated by the Association of Universities for Research in Astronomy (AURA) under a cooperative agreement with the National Science Foundation. LBNL is managed by the Regents of the University of California under contract to the U.S. Department of Energy. This project used data obtained with the Dark Energy Camera (DECam), which was constructed by the Dark Energy Survey (DES) collaboration. Funding for the DES Projects has been provided by the U.S. Department of Energy, the U.S. National Science Foundation, the Ministry of Science and Education of Spain, the Science and Technology Facilities Council of the United Kingdom, the Higher Education Funding Council for England, the National Center for Supercomputing Applications at the University of Illinois at Urbana-Champaign, the Kavli Institute of Cosmological Physics at the University of Chicago, Center for Cosmology and Astro-Particle Physics at the Ohio State University, the Mitchell Institute for Fundamental Physics and Astronomy at Texas A\&M University, Financiadora de Estudos e Projetos, Fundacao Carlos Chagas Filho de Amparo, Financiadora de Estudos e Projetos, Fundacao Carlos Chagas Filho de Amparo a Pesquisa do Estado do Rio de Janeiro, Conselho Nacional de Desenvolvimento Cientifico e Tecnologico and the Ministerio da Ciencia, Tecnologia e Inovacao, the Deutsche Forschungsgemeinschaft and the Collaborating Institutions in the Dark Energy Survey. The Collaborating Institutions are Argonne National Laboratory, the University of California at Santa Cruz, the University of Cambridge, Centro de Investigaciones Energeticas, Medioambientales y Tecnologicas-Madrid, the University of Chicago, University College London, the DES-Brazil Consortium, the University of Edinburgh, the Eidgenossische Technische Hochschule (ETH) Zurich, Fermi National Accelerator Laboratory, the University of Illinois at Urbana-Champaign, the Institut de Ciencies de l’Espai (IEEC/CSIC), the Institut de Fisica d’Altes Energies, Lawrence Berkeley National Laboratory, the Ludwig Maximilians Universitat Munchen and the associated Excellence Cluster Universe, the University of Michigan, NSF’s NOIRLab, the University of Nottingham, the Ohio State University, the University of Pennsylvania, the University of Portsmouth, SLAC National Accelerator Laboratory, Stanford University, the University of Sussex, and Texas A\&M University. BASS is a key project of the Telescope Access Program (TAP), which has been funded by the National Astronomical Observatories of China, the Chinese Academy of Sciences (the Strategic Priority Research Program “The Emergence of Cosmological Structures” Grant \# XDB09000000), and the Special Fund for Astronomy from the Ministry of Finance. The BASS is also supported by the External Cooperation Program of Chinese Academy of Sciences (Grant \# 114A11KYSB20160057), and Chinese National Natural Science Foundation (Grant \# 12120101003, \# 11433005). The Legacy Survey team makes use of data products from the Near-Earth Object Wide-field Infrared Survey Explorer (NEOWISE), which is a project of the Jet Propulsion Laboratory/California Institute of Technology. NEOWISE is funded by the National Aeronautics and Space Administration. The Legacy Surveys imaging of the DESI footprint is supported by the Director, Office of Science, Office of High Energy Physics of the U.S. Department of Energy under Contract No. DE-AC02-05CH1123, by the National Energy Research Scientific Computing Center, a DOE Office of Science User Facility under the same contract; and by the U.S. National Science Foundation, Division of Astronomical Sciences under Contract No. AST-0950945 to NOAO.

\end{acknowledgements}

\bibliographystyle{aa}
\bibliography{bib.bib}

\end{document}